\documentclass[times,twocolumn,final]{elsarticle}

\usepackage{outline}
\usepackage{pmgraph}
\usepackage[normalem]{ulem}
\usepackage[utf8]{inputenc}
\usepackage{amssymb}
\usepackage{hyperref}
\usepackage{amsmath}
\usepackage{graphicx}
\usepackage{times}
\usepackage{xcolor}
\usepackage{xspace}
\usepackage[colorinlistoftodos]{todonotes} 
\usepackage{bm} 
\usepackage{url}

\usepackage{epstopdf}
\AppendGraphicsExtensions{.tiff}
\graphicspath{{fig/}} 

\usepackage{epsfig}
\usepackage{tikz}
\usetikzlibrary{spy}
\usepackage{algpseudocode}
\usepackage{algorithm}
\usepackage{mathrsfs}

\def\QED{~\rule[-1pt]{5pt}{5pt}\par\medskip}

\long\def\comment#1{} 

\newcommand{\xmath}[1] {\ensuremath{#1}\xspace}
\newcommand{\blmath}[1] {\xmath{\bm{#1}}}

\newcommand{\I}{\blmath{I}}

\newcommand{\x}{\blmath{x}}

\newcommand{\y}{\blmath{y}}

\newcommand{\1}{\blmath{1}}

\newcommand{\xb}{{\blmath x}}
\newcommand{\yb}{{\blmath y}}
\newcommand{\zb}{{\blmath z}}

\newcommand{\Tc}{\mathcal{T}}
\newcommand{\Xc}{\mathcal{X}}
\newcommand{\Yc}{\mathcal{Y}}

\newcommand{\Ed}{{{\mathbb E}}}

\newcommand{\beq}{\begin{equation}}
\newcommand{\eeq}{\end{equation}}
\newcommand{\beqa}{\begin{eqnarray}}
\newcommand{\eeqa}{\end{eqnarray}}

\usepackage{multirow}
\usepackage{makecell}
\usepackage{booktabs}
\usepackage[flushleft]{threeparttable}
\usepackage{etoolbox}
\AtBeginEnvironment{figure}{\setlength{\intextsep}{0.5\baselineskip}}

\usepackage{medima}
\usepackage{framed,multirow}

\usepackage{latexsym}

\usepackage{url}
\usepackage{xcolor}

\usepackage{hyperref}

\definecolor{newcolor}{rgb}{.8,.349,.1}

\begin{document}

\verso{J. Huh \textit{et~al.}}

\begin{frontmatter}

\title{Tunable Image Quality Control  of 3-D Ultrasound using Switchable CycleGAN}

\author[1]{Jaeyoung Huh}
\author[1]{Shujaat Khan}
\author[2]{Sungjin Choi}
\author[2]{Dongkuk Shin}
\author[3]{Eun Sun Lee$^*$}
\author[1]{Jong Chul Ye$^*$}
\cortext[cor1]{Corresponding author:  E-mail address : seraph377@cau.ac.kr (E.S.Lee); jong.ye@kaist.ac.kr (J.C.Ye)}

\address[1]{Department of Bio and Brain Engineering, Korea Advanced Institute of Science and Technology (KAIST), Daejeon 34141, Republic of Korea}
\address[2]{System R\&D Group, Samsung Medison Co., Ltd., Seoul, Korea}
\address[3]{Department of Radiology, Chung-Ang University Hospital, 102 Heukseok-ro, Dongjak-gu, Seoul 06973, Korea}

\begin{abstract}
In contrast to  2-D ultrasound (US) for uniaxial plane imaging, a 3-D US imaging system can visualize a volume along three axial planes. This allows for a full view of the anatomy, which is useful for gynecological (GYN) and obstetrical (OB) applications. Unfortunately, the 3-D US has an inherent limitation in resolution compared to the 2-D US. In the case of 3-D US with a 3-D mechanical probe, for example, the image quality is comparable along the beam direction, but significant deterioration in image quality is often observed in the other two axial image planes. To address this, here we propose a novel unsupervised deep learning approach to improve 3-D US image quality. In particular, using {\em unmatched} high-quality 2-D US images as a reference, we trained a recently proposed switchable CycleGAN architecture so that every mapping plane in 3-D US can learn the image quality of 2-D US images. Thanks to the switchable architecture, our network can also provide real-time control of image enhancement level based on user preference, which is ideal for a user-centric scanner setup. Extensive experiments with clinical evaluation confirm that our method offers significantly improved image quality as well user-friendly flexibility.
\end{abstract}


\end{frontmatter}

 \begin{figure*}[t]
  \center
  		\vspace*{-0.3cm}
	\includegraphics[width=\textwidth]{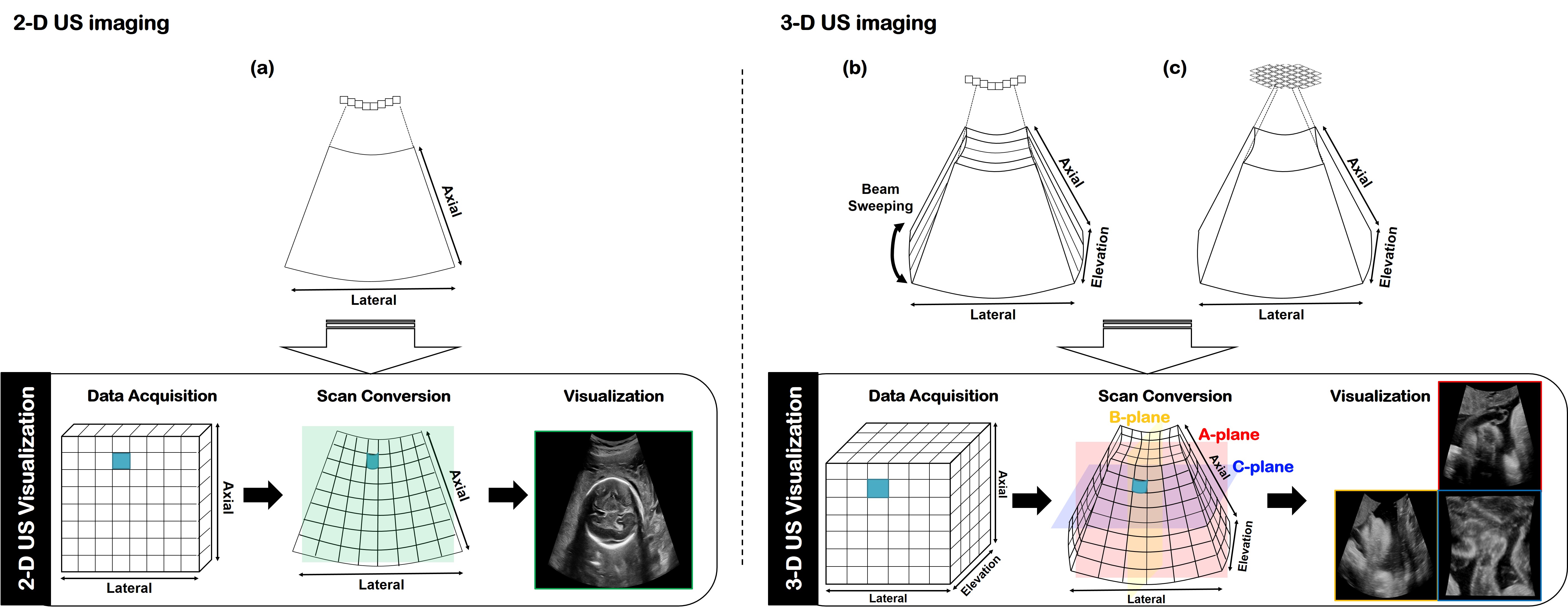}
		\vspace*{-0.7cm}
	\caption{US image generation steps. In the top row, various form of 2-D and 3-D  probes are presented. The second row presents the three step procedure of US imaging: data acquisition, scan-conversion, and visualization. (a) Convex array probe for 2-D US imaging, (b) 3-D mechanical probe  by  scanning  1-D convex array, and (c)  2-D matrix array for 3-D US imaging.}
	\label{fig:3-D}
\end{figure*}
\section{Introduction}
\label{sec:introduction}
In contrast to computed tomography (CT) and magnetic resonance imaging (MRI), ultrasound imaging (US) 
offers real-time imaging without any radiation risk.
 As such, US imaging is useful in many clinical areas, {including} gynecology (GYN) and obstetrics (OB).

Commonly used 2-{dimensional} (2-D) US systems capture images in a single plane along the direction of incidence of waves and deliver a high-quality image with the aid of a dense 1-D transducer array. On the other hand, 3-D US imaging systems can create  {3-D} images by acquiring three-axis data using mechanical scanning of 1-D transducers or 2-D array transducers (see Fig.~\ref{fig:3-D}(b)(c)).

Unfortunately, when compared to a 2-D imaging system, 3-D US has fundamental limitations in terms of its resolution. For the case 3-D US using 2-D array transducer as shown in Fig.~\ref{fig:3-D}(c), a large number of piezo-elements are necessary to obtain high-resolution 3-D volume. However, this increases the bandwidth of the RF data considerably, so that a sparse array is often used at the expense of reduced image resolution. Although 3-D US with 3-D mechanical probe using 1-D transducer scanning as shown in Fig.~\ref{fig:3-D}(b) may alleviate this, it also suffer from anisotropic resolution originated from the mechanical scanning. Specifically, while the axial-lateral plane, called the {A-plane} image, has a comparably high quality, the axial-elevation planes called {B} and the elevation-lateral planes called {C-plane} show degraded resolution. Therefore, quality improvement for 3-D US has become an important issue for ultrasound manufacturers.  
In 3-D US with a 3-D mechanical probe, which is the main imaging platform in our study,
one of the simplest approaches to address this anisotropy in the resolution
is to acquire more data along the elevation and lateral direction at the expense of the scan time. Super-resolution approaches in the 3-D volume domain may be another way to improve resolution, but they may increase the computational burden significantly, which is not appropriate for real-time imaging.

Recently, various authors have proposed image enhancement algorithms and  beamformers that use deep neural networks (\cite{yoon2019efficient,khan2020adaptive,hyun2019beamforming,nair2018deep,brickson2021reverberation,solomon2019deep,sharifzadeh2020phase,vedula2017towards,perdios2017deep,perdios2018deep,van2019deep,luchies2018deep,khan2019deep,kokil2020despeckling,luijten2020adaptive,sadeghi2021improving,zuo2021phase,huh2021unsupervised,khan2021switchable,khan2021variational,khan2020unsupervised,dietrichson2018ultrasound}). Most of the existing approaches are based on supervised learning that requires matched ground-truth data (\cite{yoon2019efficient,khan2020adaptive,hyun2019beamforming,nair2018deep,brickson2021reverberation,solomon2019deep,sharifzadeh2020phase,vedula2017towards,perdios2017deep,perdios2018deep,van2019deep,luchies2018deep,khan2019deep,kokil2020despeckling,luijten2020adaptive,sadeghi2021improving,zuo2021phase}). Unfortunately, this approach is not suitable for our 3-D US image enhancement problem as there are no matched high-resolution  3-D US images that can be used  as ground truths for supervision. Another important limitation of existing deep learning approaches is that once the neural network is trained, it is designed to produce the same level of improvement. So if another level of improvement is required, additional training is required. This is an important disadvantage that prevents their widespread acceptance in the clinical settings, as many radiologists may have different preferences for the level of image enhancement, and the common practice is to adjust the algorithm parameters according to their preference. Therefore, it would be very useful if deep learning-based image enhancement  could also provide real-time control of image quality so that radiologists can choose their preferred setup.

To address this problem, here we propose a novel 3-D ultrasound image enhancement technique that can provide real-time image enhancement and quality control based on user preferences, and that can be trained with unmatched high quality 2-D ultrasound images. In particular, our methodology is based on the recently proposed switchable CycleGAN architecture (\cite{gu2021adain,yang2021continuous}), which has been used successfully for various unsupervised learning problems. In contrast to the standard CycleGAN with two different generators (\cite{zhu2017unpaired}), the switchable CycleGAN  (\cite{gu2021adain,yang2021continuous}) has a single generator whose role can be controlled by Adaptive Instance Normalization (AdaIN) code (\cite{huang2017arbitrary}). In addition to the reduced complexity due to the common generator, one of the most important by-products of the switchable CycleGAN is its tuneability. In particular, once the network is trained, we can provide a real-time control of the level of image enhancement by simply interpolating the AdaIN codes (\cite{gu2021adain,yang2021continuous}). Therefore, this architecture successfully addresses two fundamental limitations of the existing deep learning approaches for US: lack of matched reference data and user preference-based control. Extensive experiments with clinical evaluation also confirm that the proposed method offers significantly improved image quality compared to the original 3-D scan.
   
This paper is structured as follows. Section~\ref{sec:contribution} provides necessary backgrounds and explains the
proposed method. The implementation details of the algorithm are presented in Section~\ref{sec:implementation}. Section~\ref{sec:results} provides reconstruction results using various examples, followed by a discussion in Section~\ref{sec:Discussion} and a conclusion in Section~\ref{sec:conclusion}.

\section{Main Contribution}
\label{sec:contribution}

\subsection{3-D Ultrasound Imaging Systems}
US imaging is  composed of three steps: data acquisition, scan conversion and visualization. The data acquisition step is to collect the {radio-frequency (RF)} data with a  2-D or 3-D probe. Scan conversion refers to transforming the shape of the collected RF data into a predefined shape. The visualization step creates the final image for display.

As shown in Fig. \ref{fig:3-D}(a), a 2-D system receives RF data using a 1-D array probe in a plane  that has lateral and axial axes. On the other hand, a 3-D US system  captures {3-D} information with lateral, axial and elevation axes. 
In practice, obtaining 3-D RF data can be achieved by two representative probes: 2-D array probe (\cite{yen2000sparse,huang2017review}), and 3-D mechanical probe (\cite{huang2017review,fenster2001three,fenster19963,fenster2011three}). The 2-D array transducer generates a 3-D convex waveform and then collects return echoes. On the other hand, a 3-D mechanical probe with a 1-D array transducer provides a 2-D convex waveform and sweep to acquire multiple 2-D RF data along the elevation axis. 

Although the 3-D US system provides volumetric information with a single attempt, it should be noted that with a 3-D imaging system, compared to a 2-D system, less information is captured in each plane. Specifically, it should take tremendous information in a short period of time, but there is a limitation in increasing the piezo-elements due to the transmission bandwidth. For this reason, less information is used to form the image of each slice, which causes blurring throughout the image. In particular,  3-D US systems with 3-D mechanical probe, which is the target system of our study, the elevation axis consists of more sparse data, so that the axial-elevation plane and the elevation-lateral plane have a lower quality than the axial-lateral plane. This fundamental difference inevitably leads to a deterioration in the quality of the visualization process.

Specifically, as shown in Fig. \ref{fig:3-D}, the red-colored plane, the axes of which are axial-lateral forms the {A-plane}, which is similar to the 2-D acquisition mode, but of lower quality. The plane with the axial-elevation axes, the color of which is yellow, forms the {B-plane}, whereas the plane of the elevation-lateral axes with blue color forms the {C-plane}. As mentioned before, the B-plane and C-plane are reconstructed from a smaller amount of RF data compared to A-plane, so they lose some detail and suffer from some blurring artifact.

\subsection{Unsupervised 3-D US Enhancement Using 2-D US data}

\begin{figure}[!hbt]
	\center
	\includegraphics[width=\columnwidth]{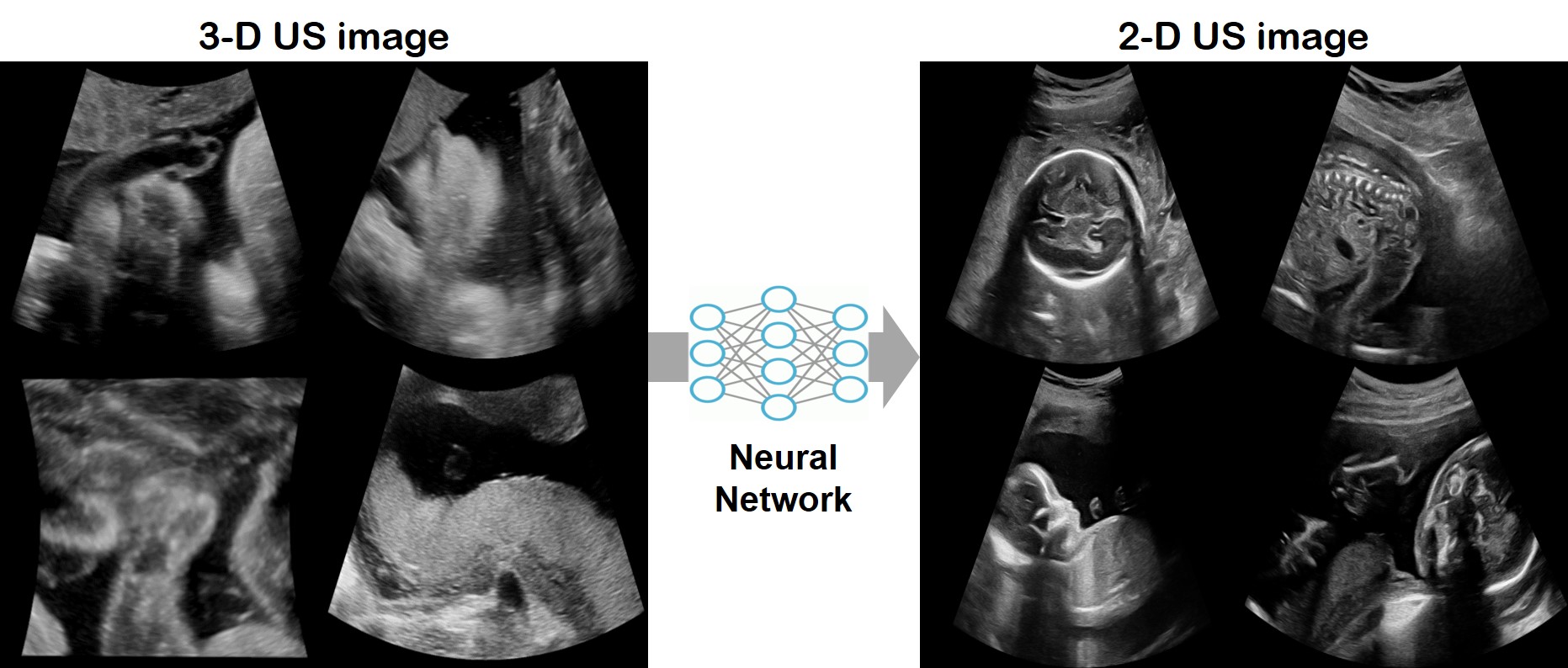}
	\vspace*{-0.7cm}
	\caption{Unsupervised learning to learn 2-D US image quality using  A-, B-, and C- plane images from 3-D US.}
	\label{fig:concept}
\end{figure}

Due to the aforementioned fundamental limitations of 3-D scanning, image quality degradation compared to 2-D US is  inevitable. To address this problem, we are interested in improving 3-D US image quality by learning image quality of 2-D US images. As shown in Fig.~\ref{fig:concept}, the neural network input is a 2-D image from any plane in 3-D volume, such as A-, B-, or C- planes, and our goal is to improve their image quality by learning image quality from 2-D US images.

 Unfortunately,  obtaining image from same area under the same condition using  2-D and 3-D probes is impossible, therefore supervised training with the paired dataset  is not feasible. For the unsupervised training with unpaired data set, one of the most powerful tools is the CycleGAN architecture (\cite{zhu2017unpaired}). Originally proposed for style-transfer for computer vision applications, CycleGAN has been successfully used for various bio-medical imaging fields such as CT (\cite{kang2019cycle,li2019low,gu2021adain,yang2021continuous,gu2021cyclegan}), MRI (\cite{oh2020unpaired,chung2021two,cha2020unpaired,modanwal2020mri,modanwal2021normalization}),  ultrasound (\cite{khan2021variational,huh2021unsupervised}), and optics (\cite{chung2021missing,lim2020cyclegan,lee2019three}).
Moreover, it was shown that CycleGAN  learns the optimal transport map that can transport one probability distribution to another (\cite{sim2020optimal2}). Accordingly, when applied to our problem, CycleGAN can be trained to transport the distribution of A-, B-, and C-plane images from 3-D US to that of the high-resolution 2-D US images.

\begin{figure*}[!t]
	\center
	\includegraphics[width=0.75\textwidth]{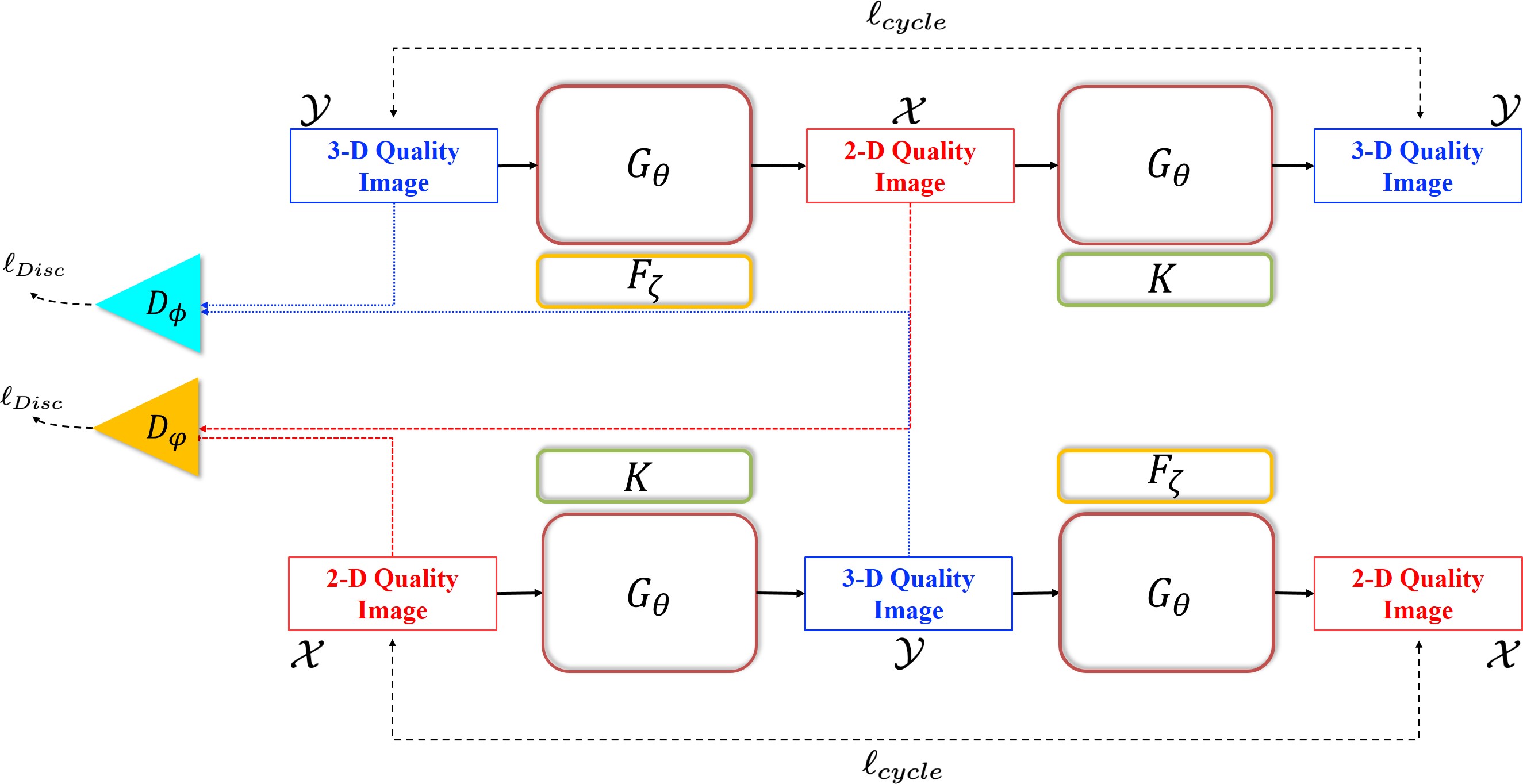}
	\vspace*{-0.3cm}
	\caption{Overview of the switchable CycleGAN for 3-D ultrasound image enhancement. The generator $G_\theta$ generates the target 2-D quality image
	from A-, B-, and C- plane images from 3-D US using the AdaIN code created from the AdaIN code generator, $F_\zeta$. The improved images are then
	 transformed into original image quality through $G_\theta$ with constant AdaIN code $K$. The discriminators $D_\phi, D_\varphi$ determine whether the image is real or fake.}
	\label{fig:scheme}
\end{figure*} 

Furthermore, it was recently shown that CycleGAN can be synergistically combined
with Adaptive Instance Normalization (AdaIN) (\cite{huang2017arbitrary}) so that
only a single generator can be used as a forward and inverse generator (\cite{gu2021adain,yang2021continuous}). By combining with a light weighted AdaIN code generator, this architecture not only reduces the number of trainable parameters to make the training more stable, but also enables the tuneability which enables the inference-time control of the neural network output as demonstrated in low-dose CT denoising (\cite{gu2021adain}) and CT kernel conversion (\cite{yang2021continuous}).
In the following, we provide more details of our switchable CycleGAN architecture.

\subsection{Switchable Generator using AdaIN Layers}

Fig. \ref{fig:scheme} illustrates the proposed switchable CycleGAN network for 3-D US image enhancement. The network architecture consists of a single generator $G_\theta$  and an AdaIN code generator, $F_\zeta$, which are parameterized by $\theta$ and $\zeta$, respectively.
The main generator $G_\theta$ in the upper left part converts US images in 3-D quality into a 2-D quality domain according to the AdaIN code created from $F_\zeta$. The right part generator returns the fake 2-D quality images to the original 3-D US quality domain, using a constant AdaIN code $K$ that corresponds to the instance normalization. There are also two discriminators $D_\phi$ and $D_\varphi$, parameterized by $\phi$ and $\varphi$, respectively, as in the conventional CycleGAN.

The main mathematical origin of replacing the two generators with a single baseline network  with two specific AdaIN codes is from the optimal transport theory (\cite{villani2009optimal,peyre2019computational}). Specifically, it was shown that the AdaIN transform corresponds to the optimal transport between two uncorrelated Gaussian probability distribution (\cite{mroueh2020wasserstein}).
More specifically,  the AdaIN layers take a mean vector  $c_{\mu}$ and a standard deviation vector   $c_{\sigma}$ as input:
\begin{align}\label{eq:AdaIN_code}
\Tc(\zb,F_\zeta(c))  = \frac{c_{\sigma}\1 }{\sigma(\zb) }\left(\zb -\mu(\zb) \1\right) +c_{\mu}\1,\quad 
\end{align}
with
\begin{align}\label{eq:AdaIN_code_generation}
\left(c_{\mu}, c_{\sigma}\right)  = F_\zeta(c) 
\end{align}
where $F_\zeta$ is the AdaIN code generator, and the $c$ is its input vector.  Then, this corresponds to the optimal transport from the distribution $\zb\sim{\mathcal N}(\mu(\zb),\sigma^2(\zb)\I)$ to another distribution ${\mathcal N}(c_\mu\1,c_\sigma^2\I)$, where ${\mathcal N}(\cdot,\cdot)$ refers to the Gaussian distribution  (\cite{mroueh2020wasserstein}).

Accordingly, for the 3-D US quality images  to high-quality 2-D US images, we use an auto-encoder as a baseline network and then use AdaIN transform to transport the auto-encoder features to the 2-D US image features. Similarly, for the translation from high-quality 2-D US images to 3-D US quality images, the auto-encoder features are then transported to the 3-D US quality image features using another AdaIN transform.

In particular, during the 3-D US enhancement task, the mean and variance vectors for the AdaIN layers are from AdaIN code generators, whereas for the generator to converting 2-D quality images to 3-D quality images the AdaIN code is set to a constant vector $K$ composed of the mean and standard deviation vector  set to zero and one, respectively:
\begin{align}
K = (0,1)
\end{align}
At the inference phase, we use the trained auto-encoder with the trained AdaIN code generator as the 3-D US image enhancement network so that it gives us an advantage of controlling image quality at the inference phase by interpolating the generated AdaIN code and the constant vector $K$:
\begin{align}
    H(c, \alpha) = (1-\alpha)K + \alpha F_\zeta(c),\quad 0\leq \alpha \leq 1
\end{align}
where $H(\cdot,\cdot)$ is an interpolated AdaIN code between $K$ and $F_\zeta(c)$. 
While the large $\alpha$ value implies more transition to the 2-D quality features, the small $\alpha$ more preserves 3-D quality features. 
Accordingly, the intermediate level image enhancement results 
 along the continuous optimal transport path between the domain $\Yc$ and $\Xc$ are obtained by (\cite{gu2021adain,yang2021continuous})
\begin{align}
    \xb_\alpha = G_\theta(\yb, H(c,\alpha)).
\end{align}
depending on user-preference, where $\yb$ denotes the 3-D US quality input images
from A-, B-, or C- planes.

\begin{figure*}[!ht]
	\center
	\includegraphics[width=0.8\textwidth]{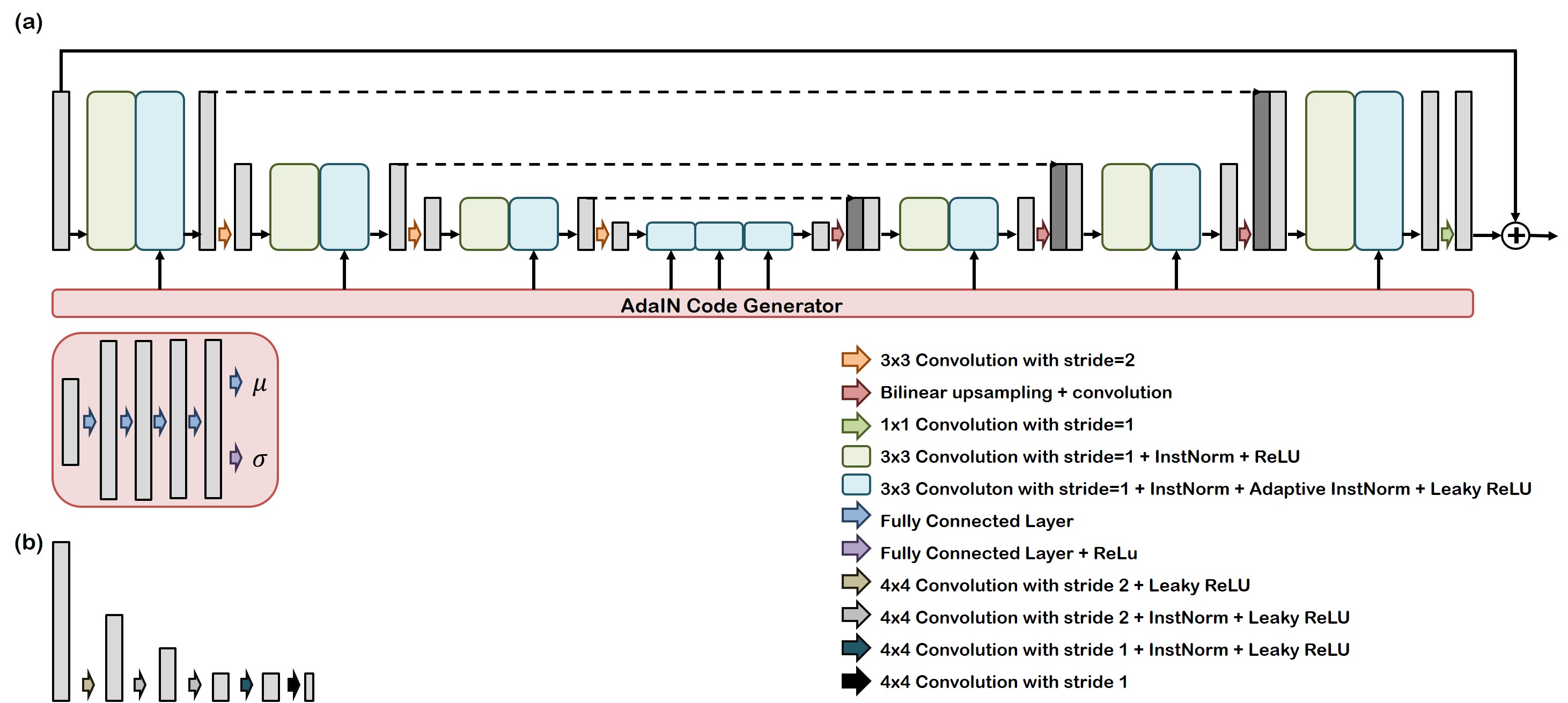}
	\caption{The proposed network architecture. (a) The proposed generator architecture. (b) The proposed discriminator architecture.}
	\label{fig:network}
\end{figure*} 

\subsection{Network Training}
The network training was  similar to the conventional CycleGAN. 
Specifically, the total loss function is represented by
 \begin{align}\label{eq:final}
\min_{\theta,\zeta}\max_{\phi,\varphi} \ell_{total}(\theta,\zeta;\phi,\varphi),
\end{align}
 where
 \begin{align} \label{eq:total}
 \begin{split}
\ell_{total}(\theta,\zeta;\phi,\varphi)& := \lambda_{cyc} \ell_{cycle}(\theta,\zeta) \\ &+\ell_{Disc}(\theta,\zeta;\phi,\varphi) + \lambda_{iden} \ell_{iden}(\theta,\zeta)
\end{split}
\end{align}
where the cycle-consistency term is given by
\begin{align}
\begin{split}
    \ell_{cycle}(\theta,\zeta)  = &\Ed_{\y \sim P_y} [||\y-G_\theta(G_\theta(\y,F_\zeta(c)),K)||_1] \\
    &+\Ed_{\x \sim P_x} [\|\x- G_\theta(G_\theta(\x,K),F_\zeta(c))\|_1],
\end{split}
\end{align}
and the discriminator term is composed of LSGAN loss (\cite{mao2017least}):
\begin{align}
    \begin{split}
&\ell_{Disc}(\theta,\zeta;\phi,\varphi)  \\
= & \Ed_{\y \sim P_y} [||D_\phi(\y)||_2^2] + \Ed_{\x \sim P_x} [||1-D_\phi(G_\theta(\x, K))||_2^2]  \\ 
 & + \Ed_{\x \sim P_x} [||D_\varphi(\x)||_2^2] + \Ed_{\y \sim P_y} [||1-D_\varphi(G_\theta(\y,F_\zeta(c))||_2^2].
    \end{split}
\end{align}
and the identity term is given by
\begin{align}
\begin{split}
    \ell_{iden}(\theta,\zeta)  = & \Ed_{\y \sim P_y} [||\y- G_\theta(\y,K)||_1] \\
&+ \Ed_{\x \sim P_x} [\|\x-G_\theta(\x,F_\zeta(c))\|_1] .
\end{split}
\end{align}
During the training,  the  discriminator $D_\varphi$ tries to differentiate between the real image $\x$ and the generated image $G_\theta(\y,F_\zeta(c))$ to confirm whether it is real or fake. Similarly, $D_\phi$ tries to find real or fake values between $\y$ and $ G_\theta(\x,K)$. It should be noted  that the proposed method has the inter-dependent $G_\theta$ and $F_\zeta$ so that they should be trained at the same time. 

\begin{table*}[h!]
	\centering
	\caption{Image Quality Scoring Criteria}
	\vspace*{0.1cm}
	\resizebox{0.8\textwidth}{!}{
		\begin{tabular}{c||c|c|c||c|c}
			\hline 
			Score & Conventional US artifact & Contrast & Overall quality & Score & Blurring by denoising \\
			\hline \hline
		    \multirow{2}{*}{5} & \multirow{2}{*}{Subtle}& \multirow{2}{*}{Excellent} & \multirow{2}{*}{Excellent} &  \multirow{4}{*}{2} &  \multirow{4}{*}{Substantial (negative effect on image quality)} \\ 
		    &  &  &  & &  \\ \cline{1-4}
            \multirow{2}{*}{4} & \multirow{2}{*}{Mild}& \multirow{2}{*}{Good} & \multirow{2}{*}{Good} &                     &  \multirow{2}{*}{} \\ 
		    &  &  &  & &  \\ \hline
		    \multirow{2}{*}{3} & \multirow{2}{*}{Moderate}& \multirow{2}{*}{Moderate} & \multirow{2}{*}{Moderate} &  \multirow{4}{*}{1} &  \multirow{4}{*}{Minimal} \\ 
		    &  &  &  & &  \\ \cline{1-4}
		    \multirow{2}{*}{2} & \multirow{2}{*}{Severe}& \multirow{2}{*}{Poor} & \multirow{2}{*}{Poor} &                     &  \multirow{2}{*}{} \\ 
		    &  &  &  & &  \\ \hline
		    \multirow{2}{*}{1} & \multirow{2}{*}{Non-diagnostic}& \multirow{2}{*}{Non-diagnostic} & \multirow{2}{*}{Non-diagnostic} &  \multirow{2}{*}{0} &  \multirow{2}{*}{None} \\ 
		    &  &  &  & &  \\ \hline
		\end{tabular}
	}
	\vspace*{-0.5cm}
	\label{table:scoring_criteria}
\end{table*}

\section{Implementation Details}
\label{sec:implementation}

\subsection {Network Architecture}

   As shown in Fig.~\ref{fig:network}(a), we employed the simple U-Net architecture with residual learning. It is composed of three down-sampling and up-sampling steps. The convolution with stride 2 is utilized for down-sampling operation. The bilinear-upsampling with a single convolution layer is used for up-sampling operation. In the green box, we used the instance normalization instead of the batch normalization. The AdaIN operation is applied in each step with instance normalization where marked as blue box. 
   
      The discriminator is shown in Fig. \ref{fig:network}(b). We used PatchGAN which classifies the local image whether it is real or fake. It is simply composed of three convolution sets of 4$\times$4 kernel size with stride 2 and LeakyReLU. The fourth convolution is same as the previous three convolution steps except for the stride size. In the last layer, there is only 4$\times$4 convolution with stride 1.
   
      The AdaIN code generator,  $F_\zeta$, receives constant one vector of $1 \times 128$ dimension as input, which is converted into AdaIN code that has the statistical information of 2-D quality domain. The AdaIN code generator is simply composed of four fully connected layers. It has multi-head structure: one convolution layer for mean and the other layer for variance. There is an additional ReLU step for the variance to avoid the negative value. While the four fully connected layers share their parameters, the last multi-head layer has its own parameters to generate a value according to the number of channels in each step. In this study, the output of AdaIN code generator is 9 pairs of mean and variance.

\subsection {Dataset}

{The datasets are composed of US image volumes from gynecology (n=208) and obstetrics (n=115) for routine clinical check. All subjects underwent US examinations through WS80A, HERA W10 US machines (Samsung Medison, Seoul, Korea) using CV1-8AD(1$\sim$8 MHz), CV1-8A(1$\sim$8MHz), V4-8(4$\sim$8MHz) US probes for GYN examinations and V5-9(5$\sim$9MHz), EV3-10B(3$\sim$10MHz) US probes for OB examinations, respectively.}

{Because OB data showed severe heterogeneity in image characteristics according to the fetal age, position and body part, we decided to use only the GYN data as training input data and all OB volumes as test data.}
We divided a total of 208 GYN volumes into 100 for training, 15 for validation and 93 for test. In each volume, we chose the plane along the three axes around the center of the volume. A total of 18,424 images were used for training input and 630 images for validation. In the case of OB, we used 13 volumes with 567 images for validation and 102 volumes for test. In particular, we tried to acquire different types of images in order to cover the broad range of structural properties for the use of GYN training set and OB validation dataset. 
{For example, we used images of bilateral ovaries and uterus for GYN training set. In terms of OB validation set, the whole body images of the first trimester were used, whereas fetal abdomen, extremities, head and spinal images were selected in case of the second trimester.} However, to fair comparison, we used only ovary region for GYN and head, face region for OB to evaluate the method.

The target datasets consist of 2-D US images acquired from WS80A, HERA W10 using C2-6(2$\sim$6MHz), CA1-7A(1$\sim$7MHz). Only OB datasets are used as these are more suitable for the purpose of this research such as artifact suppression and sharp boundary. The images are captured by various structures and status. We selected 14,995 images for the target dataset.

  \begin{figure*}[!hbt]
    \center
  	\includegraphics[width=0.9\textwidth]{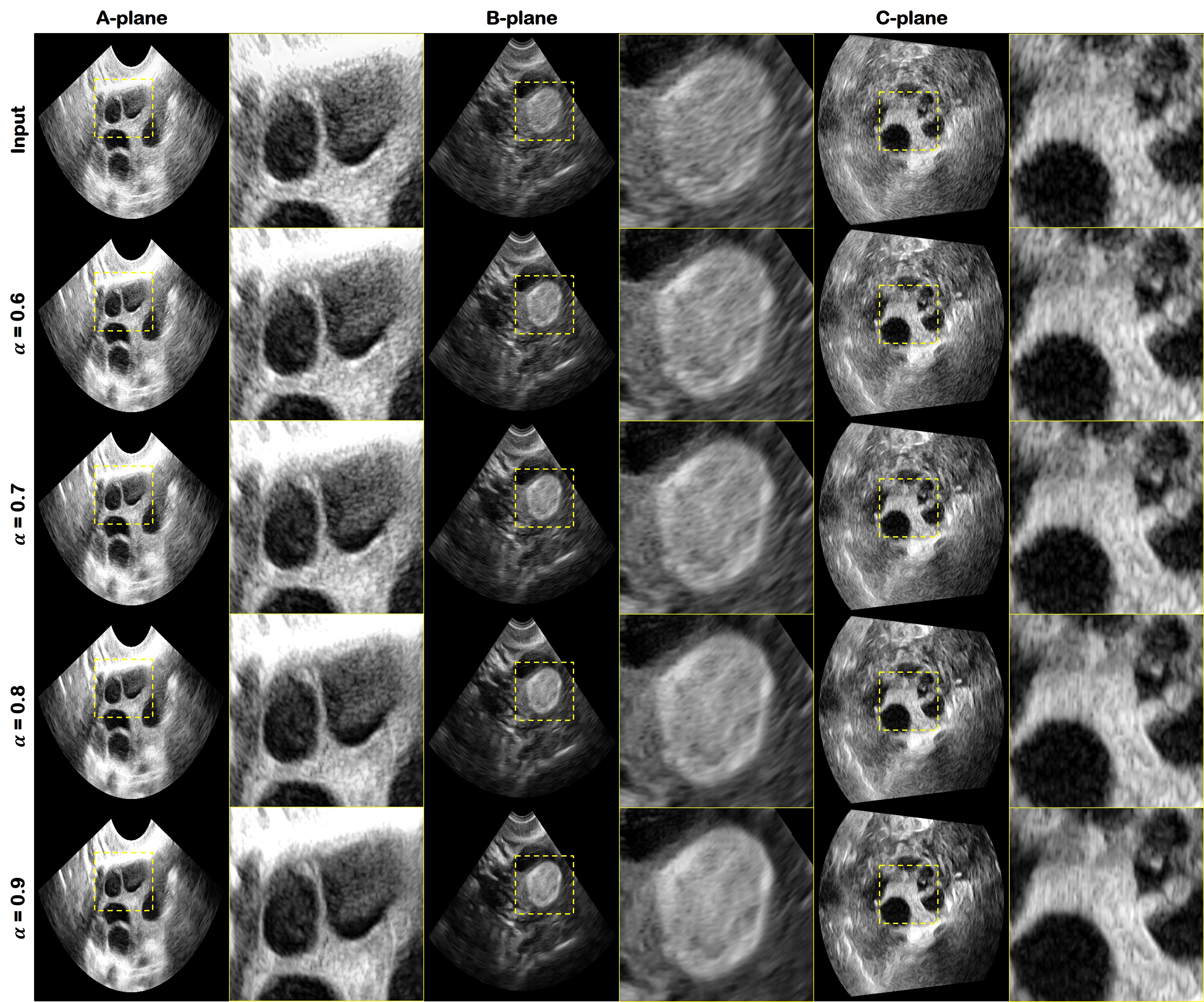}
	\vspace*{-0.2cm}
  	\caption{Results of the proposed method for gynecology test set. The first row is the input image. The result image using the proposed method is displayed from the second to the last row according to the $\alpha$ value variation. The yellow box is the magnified region.}
  	\label{fig:result_gyn}
  \end{figure*}

\begin{figure*}[h!]
  \center
	\includegraphics[width=0.95\textwidth]{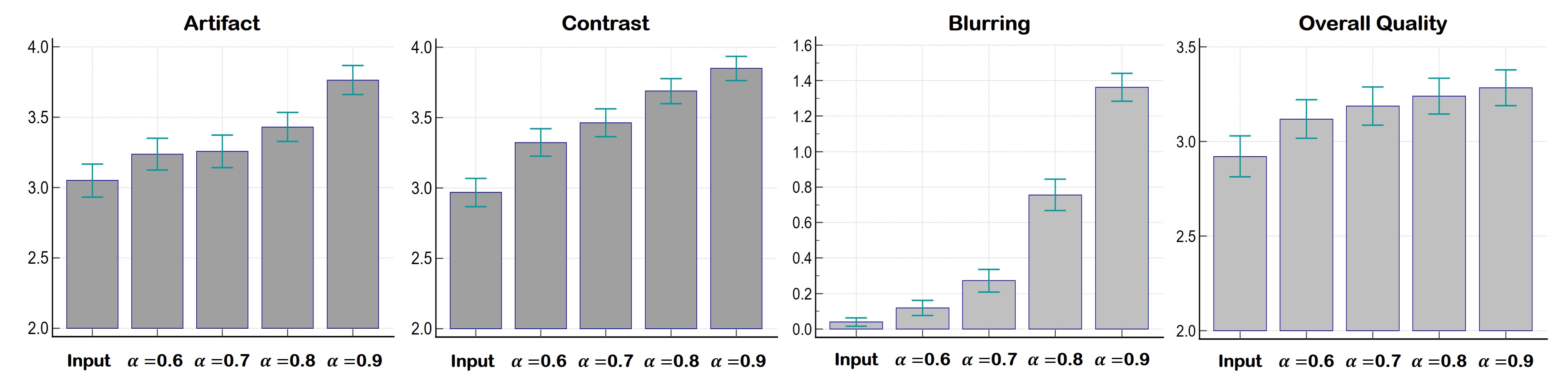}
		\vspace*{-0.3cm}
	\caption{Clinical evaluation results. Each statistic is calculated using the Friedman test with Bonferroni correction. The first and second graphs are artifact and contrast values. The third and fourth graphs are the blurring effect and the overall quality score. Each bar in the chart shows the input and results of the proposed method from $\alpha = 0.6$ to $\alpha = 0.9$.}
	\label{fig:clinical}
\end{figure*}

\begin{table*}[h!]
    \caption{Wilcoxon paired test result}
    \centering
	\resizebox{\textwidth}{!}{
		\begin{tabular}{c||c|c|c|c|c|c|c|c|c|c}
			\hline 
			Score Category & a \& b & a \& c & a \& d & a \& e & b \& c & b \& d & b \& e & c \& d & c \& e & d \& e \\ \hline
			Artifact & \textcolor{red}{0.0471} & \textcolor{red}{0.0290} & \textcolor{red}{$<$ 0.0001} & \textcolor{red}{$<$ 0.0001} & 0.8233 & \textcolor{red}{0.0091} & \textcolor{red}{$<$ 0.0001} & \textcolor{red}{0.0294} & \textcolor{red}{$<$ 0.0001} & \textcolor{red}{$<$ 0.0001} \\ \hline
			Contrast & \textcolor{red}{$<$ 0.0001} & \textcolor{red}{$<$ 0.0001} & \textcolor{red}{$<$ 0.0001} & \textcolor{red}{$<$ 0.0001} & \textcolor{red}{0.0468} & \textcolor{red}{$<$ 0.0001} & \textcolor{red}{$<$ 0.0001} & \textcolor{red}{0.0017} & \textcolor{red}{$<$ 0.0001} & \textcolor{red}{0.0139} \\ \hline
			Blurring & \textcolor{red}{0.0046} & \textcolor{red}{$<$ 0.0001} & \textcolor{red}{$<$ 0.0001} & \textcolor{red}{$<$ 0.0001} & \textcolor{red}{0.0003} & \textcolor{red}{$<$ 0.0001} & \textcolor{red}{$<$ 0.0001} & \textcolor{red}{$<$ 0.0001} & \textcolor{red}{$<$ 0.0001} & \textcolor{red}{$<$ 0.0001} \\ \hline
			Overall Quality & \textcolor{red}{0.0214} & \textcolor{red}{0.0011} & \textcolor{red}{$<$ 0.0001} & \textcolor{red}{$<$ 0.0001} & 0.3854 & 0.1223 & \textcolor{red}{0.0191} & 0.4081 & 0.2319 & 0.6175 \\ \hline
		\end{tabular}
	}
	\begin{tablenotes}
      \small
      \item a : Input, b : $\alpha=0.6$, c : $\alpha=0.7$, d : $\alpha=0.8$, e : $\alpha=0.9$.
    \end{tablenotes}
	\label{table:clinical}
\end{table*}

\begin{figure*}[h!]
  \center
	\includegraphics[width=0.85\textwidth]{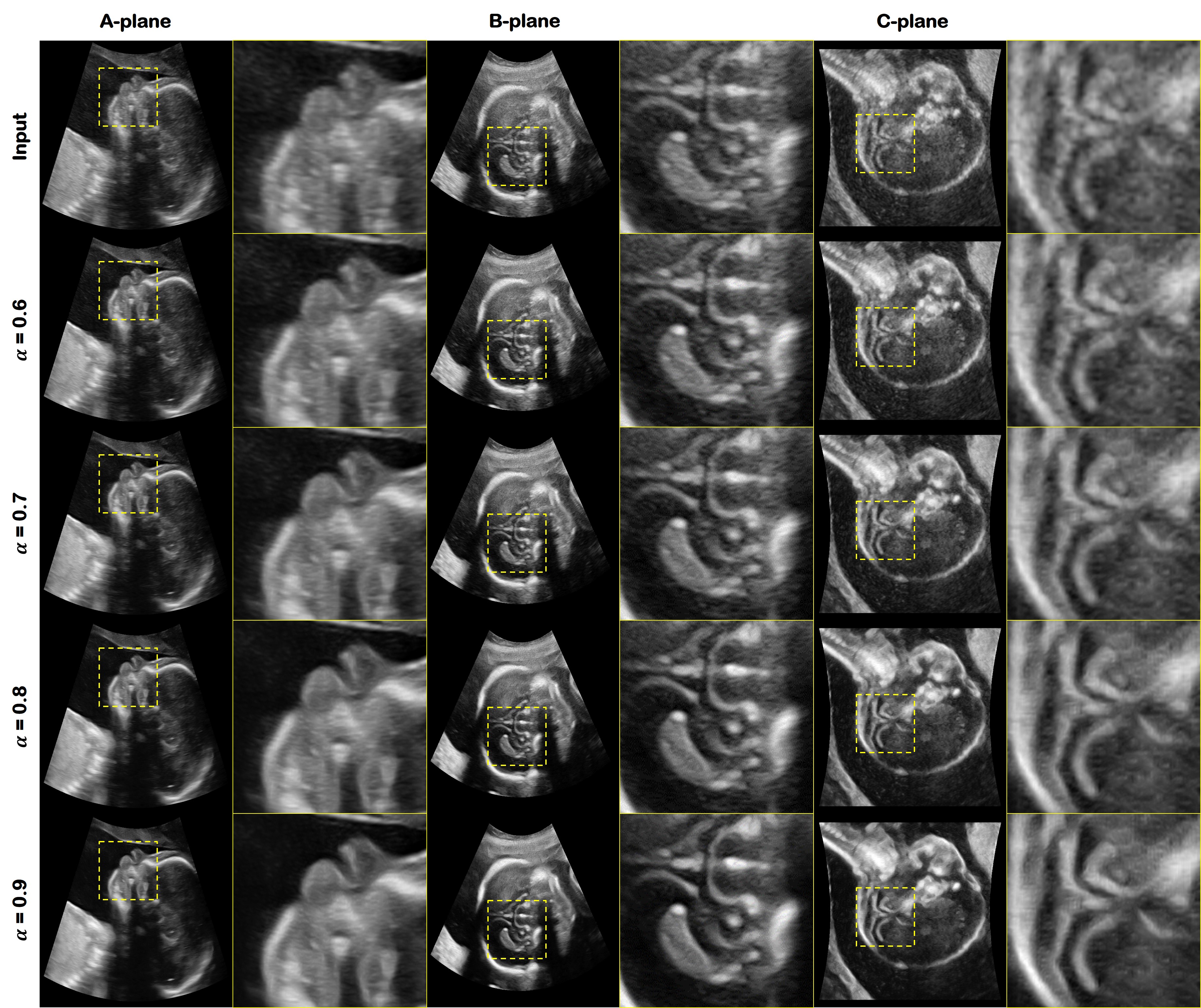}
		\vspace*{-0.3cm}
	\caption{Results of the proposed method using obstetric test set. The first row is the input image. The result images using the proposed method  are displayed from the second to the last row as a $\alpha$ value change. The yellow box is the magnified region.}
	\label{fig:result_ob}
\end{figure*}

 \begin{figure*}[h!]
  \center
  		\vspace*{-0.3cm}
	\includegraphics[width=0.8\textwidth]{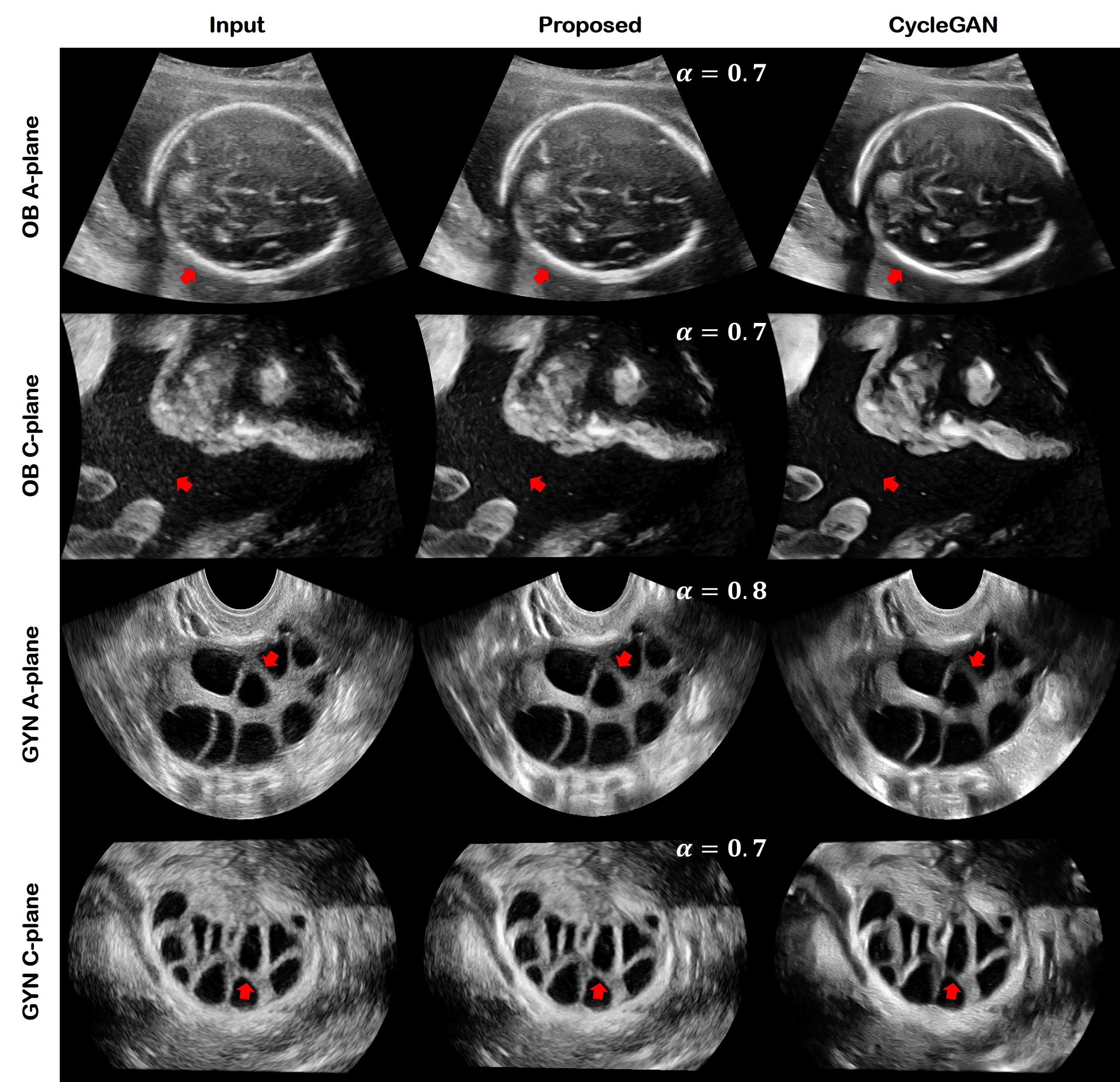}
		\vspace*{-0.3cm}
	\caption{The comparison result between the proposed method and standard CycleGAN. The first and second rows are obstetric A- and C- planes. The third and fourth rows are gynecology A- and C- planes. The first columns denotes the input image. The second and third columns are the result from the proposed method and the CycleGAN, respectively. The differences seen at the red arrows are discussed in the main text.}
	\label{fig:cyclegan}
\end{figure*}

\subsection {Training details}
To train the network, we used the Adam optimizer with $\beta=0.5$. The learning rate was started from 1e-4 and decreased linearly after 10 epochs. We set the total epoch as 50 and selected the 15 epoch which show the best result. We implemented with patch-processing whose size is 256 and the batch size was 1. To avoid the over-fitting problem, we used data augmentation technique like flipping, and rotating. The hyper-parameters were set as 10, 5 for $\lambda_{cyc}$ and $\lambda_{iden}$, respectively.  We implemented all work on the Python 3.6.12 with Tensorflow 1.14.0 and MATLAB 2017a. We trained the network with the NVIDIA GeForce GTX 1080 Ti GPU.

\subsection{Evaluation metric}
A 15-year experienced board-certified abdominal radiologist (ESL) evaluated the result of the proposed method using a test set in a blinded manner. Note that the test set was completely invisible during training. We acquired 93 volumes for quality assessment, which confined to the ovaries for the consistency of evaluation regardless of subjects' factor as possible. Evaluation statistics was implemented for input and result from $\alpha = 0.6, 0.7, 0.8, 0.9$. 

There are four categories for assessing image quality: 1) conventional US artifacts, including reverberation, side-lobes, beam thickness and etc; 2) image contrast, for example, inter-organ contrast (ovaries and adjacent structures) and intra-organ contrast (focal lesion contrast within ovaries); 3) blurring by denoising process, which means blurred textures and boundaries frequently seen after usual noise reduction; 4) overall image quality, determined subjectively. The simple evaluation criteria and scoring system are listed in the Table. \ref{table:scoring_criteria}. All statistical analysis were carried out with MedCalc version 20.015 (MedCalc software).

\section{Experimental Results}
\label{sec:results}

\subsection {Gynecology Results}
To validate our proposed method, we applied our algorithms to real gynecology (GYN) clinical data sets. The results are shown in Fig. \ref{fig:result_gyn}. The first row shows the input image. From the second to the last row, each row refers to the outputs of the proposed network with different $\alpha$ values. The left column is the A-plane, which is cut along the elevation axis, the second and third are the B-plane and the C-plane, respectively, which are cut along the lateral axis and axial axis, respectively. We have presented the images from $\alpha = 0.6$ to $\alpha = 0.9$, the change being clearly visible.

As shown in the Fig. \ref{fig:result_gyn}, the input image is gradually changed to the target style depending on $\alpha$, which can be again verified from the magnified area. The follicle boundaries are not clear from the input image. However, the proposed method connects the structures smoothly. The contrast is increased so that the follicle borders can be clearly seen. In addition, tearing artifacts in the B-plane is well suppressed and their resolution appears to be increased. The side lobe artifacts, which is one of the basic artifact in US image, are clearly visible in the input image. On the other hand, the proposed method has suppressed it and its boundaries become sharp and easily discernible as the value of $\alpha$ is increased.

\subsection {Computational Time}
Our method is based on a deep neural network, so that after training the network, the improved images can be generated in real time. The average computation time of our method is 0.139 seconds when it reconstructs the $ 968 \times 968$  image. The time is only calculated with GPU NVIDIA GeForce GTX 1080 Ti. We obtained the mean value of the computing time with 270 test images, half of which consist of GYN and half of OB.

\subsection {Clinical Evaluation}
A professional abdominal radiologist  evaluated the proposed method according to the criteria of image analysis. We took about 15 images of the A-, B- and C- planes around the ovarian region. It was processed with the proposed method using $\alpha = 0.6,0.7,0.8,0.9$. Then we randomly shuffled the set. We used 93 volumes for clinical test, so there were 1395 mixed sets for clinical evaluation.

We implemented the Friedman test with the Bonferroni correction (\cite{friedman1937use,bonferroni1936teoria}). The evaluation results are shown in Fig. \ref{fig:clinical}. The first and second graphs show the artifact and contrast rating. The third and fourth graphs represent degree of blurring and the overall quality score. According to artifact and contrast graphs, it can be easily seen that the higher the value of $\alpha$, the higher the score. In particular, the entire proposed method has a high score compared to the input image. What should be noted here is that even though blurring effect occurs, the overall image quality from the proposed method overwhelms the input images.

To check how meaningful the statistical information is, we implemented a $p$-test with Wilcoxon paired test, which is shown in the Table. \ref{table:clinical} (\cite{conover1999practical}). Each a, b, c, d, e denotes the input and from $\alpha = 0.6$ to $\alpha = 0.9$. Each row is the scoring category. We calculated a $p$-value that shows how significant the statistical differences are. The part marked in red is the statistically significant one. It should be noted that all the results of the proposed method in overall quality categories show a significant difference comparing with the input image.

\subsection{Obstetric Results}
We tested our algorithm with the OB images that were not used in network training. The goal was to demonstrate that our method can be robustly applied to a wide range of datasets.

In Fig. \ref{fig:result_ob}, we have visualized the head bone of the fetus. Each row denotes the input and its output of the proposed method with different $\alpha$ values. Each column shows the A-, B- and C- planes. As shown in Fig. \ref{fig:result_ob}, the input image has ambiguous boundaries, in particular its structure in the head bone is not easy to see. However, the result of the proposed method is that the internal structure is more visible than the input image. In addition, its boundary becomes sharp as the value of $\alpha$ is increased. In particular, the side-lobe artifact is well suppressed according to the $\alpha$ variation.

\section{Discussion}
\label{sec:Discussion}

\subsection{Comparison with conventional CycleGAN}
The proposed method is based on the CycleGAN. However, one of the big problems with applying traditional CycleGAN to this task is that the level of translation in 2-D quality cannot be controlled as desired. This means that once the network is trained, the input image is translated into 2-D quality directly. Unfortunately, there is a lot of concern that too much changes in image quality makes sense of disparity for most radiologists. It also leads to unacceptable changes in the image that hinder an accurate diagnosis. For this reason, we believe that our {Switchable CycleGAN} is very useful.

Specifically, the comparison results are shown in Fig. \ref{fig:cyclegan}. We visualized OB A-, C- planes and GYN A-, C- planes. The first column is the input image and the second, third column are the results of the proposed method and CycleGAN. Both the proposed method and the CycleGAN results show an improvement in image quality such as noise reduction and contrast enhancement. However, as can be seen in the first row where the red arrow is pointing, the bone is emphasized too much by CycleGAN. It leads to misinterpretations in the case of biparietal diameter assessment. In the second row, the red arrow points to the float in the amniotic fluid. It contains important information to interpret the status of the fetus. For example, the movement of the particles can be crucial criteria to diagnose the digestive system disorders. However, the CycleGAN recognized this as a noise and suppressed it. On the other hand, the proposed method retains it with high contrast, in particular it can be more visible by controlling the $\alpha$ value. In the third and fourth rows, the red arrow points to the ovarian structure, which can determine the status of the follicle. Since the 2-D quality emphasizes the contrast too much, it sometimes suppresses the weak signal. As you can see in the third and fourth rows, the structure becomes invisible in the CycleGAN result. On the contrary, the proposed method shows sharp boundaries compared to the input image.

\subsection {Spatial Control}
\begin{figure}[h!]
  \center
  		\vspace*{-0.3cm}
	\includegraphics[width=9cm]{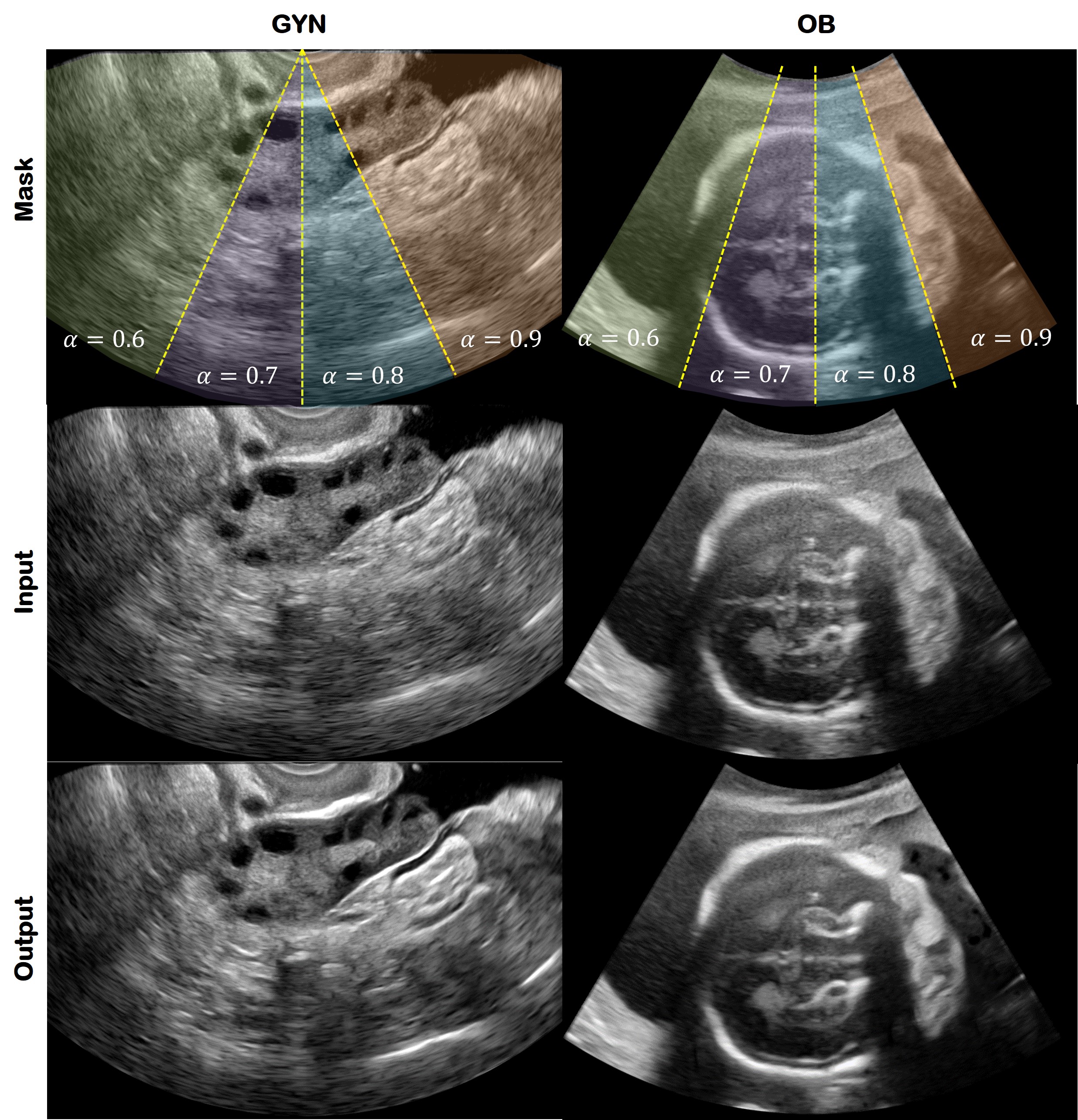}
		\vspace*{-0.5cm}
	\caption{Spatial control results. The first row is the mask that shows where and which $\alpha$ value was applied. The second and third rows are the input and output, respectively. The first column is gynecology dataset and the second column is obstetric dataset.}
	\label{fig:spatial}
\end{figure}

Another great advantage of the proposed method is that the user can select the area to be improved. Since AdaIN changes its content feature map using the style statistics generated by the AdaIN code generator, the user can control it spatially. Assume there is a mask like in the first row of Fig. \ref{fig:spatial}. There are four areas marked as green, purple, blue, and orange. Instead of applying the masking procedure four times, our method can easily generate a spatially enhanced image with a single feed-forward process. Specifically, the mask enters the network with the image. Then it is multiplied by the content feature map to normalize only the selected area of with a certain $\alpha$ value through AdaIN code interpolation. Accordingly, we can apply four different $\alpha$ values to four different regions with a single feed-forward network.

As shown in Fig. \ref{fig:spatial}, when the mask area in the top row is applied to each image, the proposed method naturally produces spatially varying quality improvement.
This provides more nature views with only enhancing area with diagnostic interest. 

\subsection {3-D Volume Rendering}
\begin{figure*}[h!]
	\center
	\includegraphics[width=\textwidth]{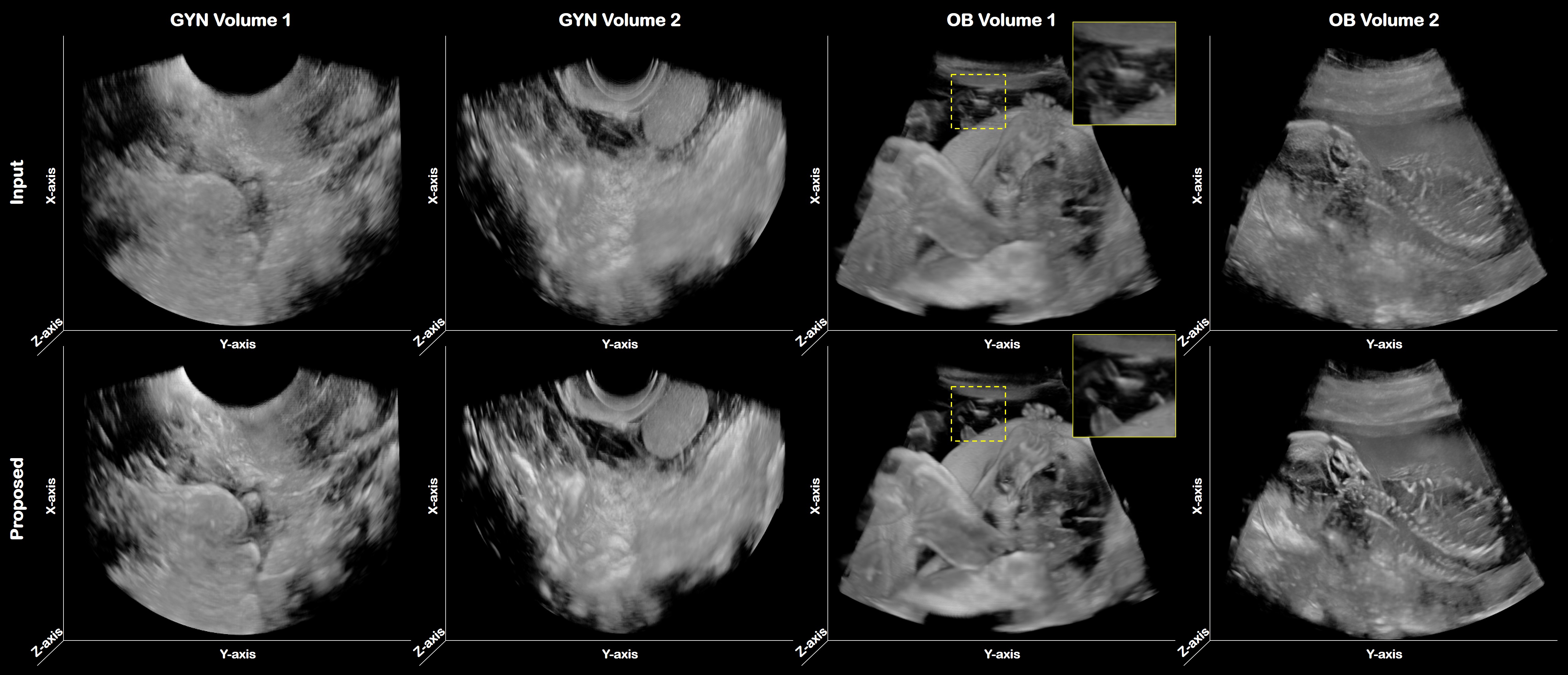}
	\vspace*{-0.5cm}
	\caption{3-D volume rendering results. The first and second row denote the input and the results from the proposed method, respectively. The first and second column are two different test gynecology volume sets, and the third and fourth column are obstetric volume test sets.}
	\label{fig:volume}
\end{figure*}

In order to verify the effect of the proposed method in the 3-D volume image, we generated a 3-D rendering view with the system {Paraview 5.10.0}. In  Fig.~ \ref{fig:volume}, we have shown two GYN and OB volume data sets. The first row is an input image and the second row is output by the proposed method. We have all generated A-plane images that are cut along the elevation axis and fed into the trained network. Then the output planes are concatenated along the elevation axis to acquire original volume data. All inputs and proposed paired volume images were visualized using the same intensity scale.

As shown in Fig. \ref{fig:volume}, the structure boundaries are clearer in the proposed volume than in the input volume. In particular, the side-lobe artifact has been suppressed and its structure is then better visible. While artifacts are shown in input volume images which is highlighted as a yellow box at the top of the image, the proposed method suppresses well. The arm and head of the fetus are more clearly visible compared to the input volume image. In the fourth column in particular, the spine of the fetus can hardly be found on the input volume image, but it is easy to identify with the proposed method.

\section{Conclusion}
\label{sec:conclusion}
Compared to the 2-D systems, the 3-D US can acquire significant rich information in a single trial. However, due to their image generation process, the 3-D US imaging suffers from deteriorated image quality. To overcome this problem, we proposed a 3-D US image enhancement method based on the Switchable CycleGAN that translates the A-, B-, and C- planes of 3-D US into high-resolution 2-D image quality. Thanks to the use of AdaIN, a single network could translate images at any planes in 3-D US into the high-quality images, and also provided continuous translation between the input and target domain.
Using extensive experiments with clinical evaluation, we have confirmed that the proposed method can improve 3-D US from low quality to high quality. In addition, we have certified the robustness of the proposed method through tests with completely invisible data sets such as obstetric datasets. Furthermore, we confirmed that the tunable nature of our method is very useful for clinical purpose by avoiding over-smoothing clinically useful features and provide a user-centric control of image quality depending on user's preference. Thanks to the outstanding performance and flexibility, we believe that the proposed method can be a useful platform for clinical 3-D US  systems.

\section*{Acknowledgement}
This work was supported by the National Research Foundation of Korea under Grant NRF-2020R1A2B5B03001980.

\bibliographystyle{model2-names.bst}\biboptions{authoryear}
\bibliography{ref}

\end{document}